\documentclass[api,prl,reprint, showpacs, preprintnumbers, amsmath, amssymb]{revtex4-1}

\usepackage{amssymb}
\usepackage[retainorgcmds]{IEEEtrantools}
\usepackage{amsmath}
\usepackage[amssymb]{SIunits} 
\usepackage{bm}
\usepackage{graphicx}
\usepackage{mathrsfs}

\newcommand{\ud}{\,\mathrm{d}}

\begin{document}
\title{Linear response theory for arbitrary periodic signals}
\date{\today}
\author{G. T. Landi}
\email{gtlandi@gmail.com}
\author{M. J. de Oliveira}
\affiliation{Instituto de F\'isica da Universidade de S\~ao Paulo,  05314-970 S\~ao Paulo, Brazil}

\begin{abstract}
We extend Kubo's Linear Response Theory (LRT) to periodic input signals with arbitrary shapes and obtain exact analytical formulas for the energy dissipated by the system for a variety of signals. These include the square and sawtooth waves, or pulsed signals such as the rectangular, sine and $\delta$-pulses.  It is shown that for a given input energy, the dissipation may be substantially augmented by exploiting different signal shapes. We also apply our results in the context of magnetic hyperthermia, where small magnetic particles are used as local heating centers in oncological treatments. 

\end{abstract}

\maketitle{}

Kubo's linear response theory \cite{Kubo1957} (LRT) has seen an immense success in the past decades. It has served as the starting point for much of the advances in statistical physics and stochastic process, while also acting as the foundation to a variety of applications \cite{Coffey2004,Scaife1971}. One of the many important breakthroughs it provides is the ability to relate the dynamic properties of the system with it's quasi-static response. This is of particular importance  in dielectric and magnetic media \cite{Coffey2004,Scaife1971}, where the response to a harmonic electric or magnetic field may be accurately related to the relaxation time of the system, provided the field amplitude is sufficiently small. Relevant examples include the electric response of polar molecules and liquid crystals \cite{Scaife1971}, or the magnetic response of small magnetic particles \cite{Brown1963,Coffey2004}. The latter, in particular, has seen renewed  interest in recent years due to it's potential application in oncological treatments via a technique known as magnetic hyperthermia \cite{Carrey2011,Pankhurst2009}. In it, with the purpose of thermally lysing tumorous cells, one exploits the heat dissipated by magnetic nanoparticles under the influence of an external high frequency magnetic field. For a harmonic stimuli, the dissipation is directly related to the imaginary part of the complex susceptibility. Thus, one may employ the LRT as a tool in the quest to optimize the heat dissipated by the particles.

The LRT, however,  is not restricted to harmonic inputs and may readily be extend to describe the system's response to any periodic stimuli expandable in a Fourier series \cite{Tolstov1962}. Albeit straightforward, to our knowledge such development has not yet been performed analytically. Clearly, it is possible to envisage several situations where these results may be of use. We will, however, focus primarily on magnetic hyperthermia. The main motivation  is that, in maximizing the dissipation of the particles, it is also necessary to maintain a  sufficiently low exciting frequency in order to  avoid the formation of eddy currents inside the patient's body \cite{Pankhurst2009}. Thus, different alternatives must be exploited, the field's shape being one possibility.

\begin{figure}[!h]
\centering
\includegraphics[width=0.45\textwidth]{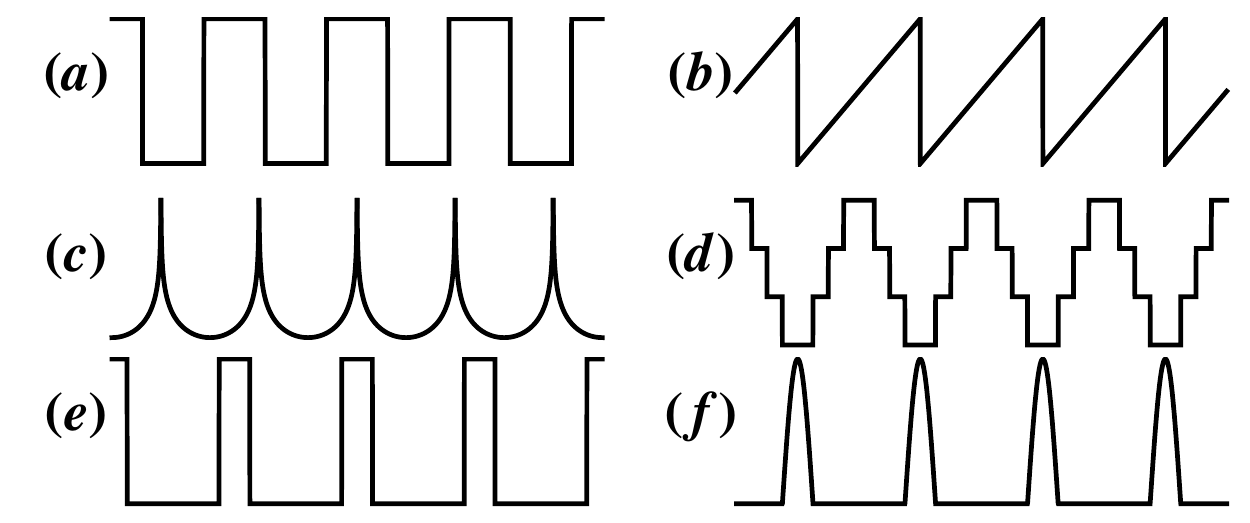}
\caption{\label{fig:col}Collection of signals studied. (a) square wave, Eq.~(\ref{eq:eta_square});  (b) sawtooth wave, Eq.~(\ref{eq:eta_sawtooth});  (c) $\eta(t) = \log[2 \cos (t/2)] $;  (d) step-ladder, Eq.~(\ref{eq:eta_stepladder});  (e) rectangular pulse, Eq.~(\ref{eq:eta_recpulse});  (f) sine pulse, Eq.~(\ref{eq:eta_sinp}).}
\end{figure}

 It is the purpose of this paper to extend the LRT to encompass arbitrary periodic signals.  First, a general formula to numerically evaluate the response is developed. Thereafter, we focus primarily  on the energy $A$ dissipated  by the system. For a variety of signals [cf. Fig.~\ref{fig:col}], we compute exact analytical formulas for $A$ in terms of $x = \omega\tau$, where $\omega$ is the exciting frequency  and  $\tau$ is the relaxation time of the system. We begin by studying simple signals such as the square and sawtooth waves. Subsequently, we turn to periodic pulsed signals and develop formulas for the  rectangular and sine pulses. From these, the $\delta$-pulse response is also readily obtained. Finally, we apply these results in the context of small magnetic particles and compare them to numerical simulations of the magnetic Langevin equation \cite{Coffey2004}.

The computations that follow are valid for any input/output system to which the LRT applies.  For concreteness, however, we shall focus on the response of a magnetic system with uniform magnetization $M(t)$, to a periodic magnetic field $H(t)$ decomposable in a Fourier series. We write $H(t) =H_0 \eta (t)$ where,
\begin{equation}\label{eq:eta_fourier}
\eta (t) = a_0 + \sum\limits_{n=1}^{\infty} \Big(a_n \cos n \omega t + b_n \sin n \omega t \Big) = \sum\limits_{n=-\infty}^{\infty}\!\!\! c_n e^{i \omega t} 
\end{equation}
For comparative purposes we normalize the energy in $\eta(t)$ to match that of a harmonic input [$\eta (t) = \cos\omega t$]; i.e., we write
\begin{equation}\label{eq:eta_norm}
\omega\!\!\!\!\!\int\limits_\text{period}\!\!\! \big| \eta(t) \big|^2 \ud t = \pi
\end{equation}
Thus, the input energy is now always equal to $\pi H_0^2$ and  the criteria for the applicability of the LRT becomes restricted exclusively to $H_0$.

Assuming that this criteria is satisfied, we expect that the magnetization will respond linearly to $H(t)$, albeit possibly with a phase lag. Whence, we write
\begin{equation}\label{eq:phase_response}
M(t) = \int\limits_{-\infty}^t H(t') \chi(t-t') \ud t,
\end{equation}
where $\chi(t)$ is a function describing the system's response. Inserting Eq.~(\ref{eq:eta_fourier}) in Eq.~(\ref{eq:phase_response}) we obtain 
\begin{equation}\label{eq:mag_fourier}
M(t) = H_0 \sum\limits_{n=-\infty}^\infty c_n \chi(n\omega) e^{in\omega t},
\end{equation}
where
\begin{IEEEeqnarray}{rCl}\label{eq:chi_kubo}
\chi(n\omega) =  \int\limits_0^\infty e^{-in\omega t} \chi(t)\ud t &=& \chi_0 \left[ 1 - i n\omega \int\limits_0^\infty C(t) e^{-in\omega t} \ud t \right]\IEEEeqnarraynumspace
\end{IEEEeqnarray}
The second equality  follows directly from the Kubo relation \cite{Kubo1957}. Here $\chi_0 = \chi(0)$ is the static susceptibility and $C(t)$ is the autocorrelation function of the system. Knowledge of these two quantities enables one to employ Eqs. (\ref{eq:eta_fourier}), (\ref{eq:mag_fourier}) and (\ref{eq:chi_kubo})  to compute the response of the system to any periodic stimuli. 


A parametric plot of $(H(t), M(t))$ yields a hysteresis loop, the area of which is precisely the average energy dissipated per cycle (which follows from the first law of thermodynamics):
$A =\int\limits\!\!H\ud M$.
Let us write $\ud M = \dot{M}(t) \ud t$. Then, since $M(t)$ is a continuous function of time [$H(t)$ need not be], we may differentiate it's Fourier series term by term \cite{Tolstov1962}. Thus,  $\dot{M}(t)$ will also be described by a Fourier series and we may use  Parseval's theorem \cite{Tolstov1962} to  write
\begin{equation}\label{eq:a2}
A = \pi H_0^2 \sum\limits_{n=1}^\infty (a_n^2+b_n^2)n [-\text{Im }\chi(n\omega)]
\end{equation} 
This formula is entirely general, valid for any autocorrelation function.

Henceforth, we specialize the calculations further  to systems described by a Fokker-Planck equation. In this case $C(t)$ is described by an infinite sum of decaying exponentials, each representing a possible relaxation mechanism of the system. Often, however, a single exponential provides the dominant contribution; i.e., $C(t) = e^{-t/\tau}$, where $\tau$ is then referred to as the relaxation time of the system. In what follows we consider only such form for $C(t)$. This assumption, however, is not restrictive given the linearity of all the equations involved: if $C(t)$ is described by more than one exponential, their contributions may simply be appended to Eq.~(\ref{eq:a2}), weighted with  proper coefficients. For $C(t) = e^{-t/\tau}$, Eq.~(\ref{eq:chi_kubo}) becomes
\begin{equation}\label{eq:chi_expo}
\chi(n\omega) = \chi_0 \frac{1}{1+in\omega \tau}
\end{equation} 
Note  that, now, $\chi(n\omega)$ is the only quantity where the frequency has a definite influence. Everywhere else we may equivalently set $\omega=1$, making the fundamental period of the signal equal to $2\pi$; thus, we henceforth take  $t\in[-\pi,\pi]$. 

Since the average input energy per cycle is $\pi H_0^2$ [cf. Eq.~(\ref{eq:eta_norm})], we may define the efficiency in \emph{converting} electromagnetic energy into thermal energy --- or, more generally, the input/output energy gain --- as $\Omega = A/\pi H_0^2$. Whence, inserting  Eq.~(\ref{eq:chi_expo}) into Eq.~(\ref{eq:a2}) we obtain:
\begin{equation}\label{eq:effi}
\Omega = \chi_0 \sum\limits_{n=1}^\infty (a_n^2 + b_n^2) \frac{n^2 x}{1+(nx)^2},\qquad x = \omega\tau
\end{equation}
This formula gives the dissipation efficiency for any system described by a single relaxation time. As mentioned, the generalization for more than one relaxation time is straightforward. It is quite remarkable that, except for $\chi_0$, all other properties of the system condense into a single variable: $x = \omega\tau$. For simplicity, we henceforth set $\chi_0=1$.

\begin{figure}[!h]
\centering
\includegraphics[width=0.5\textwidth]{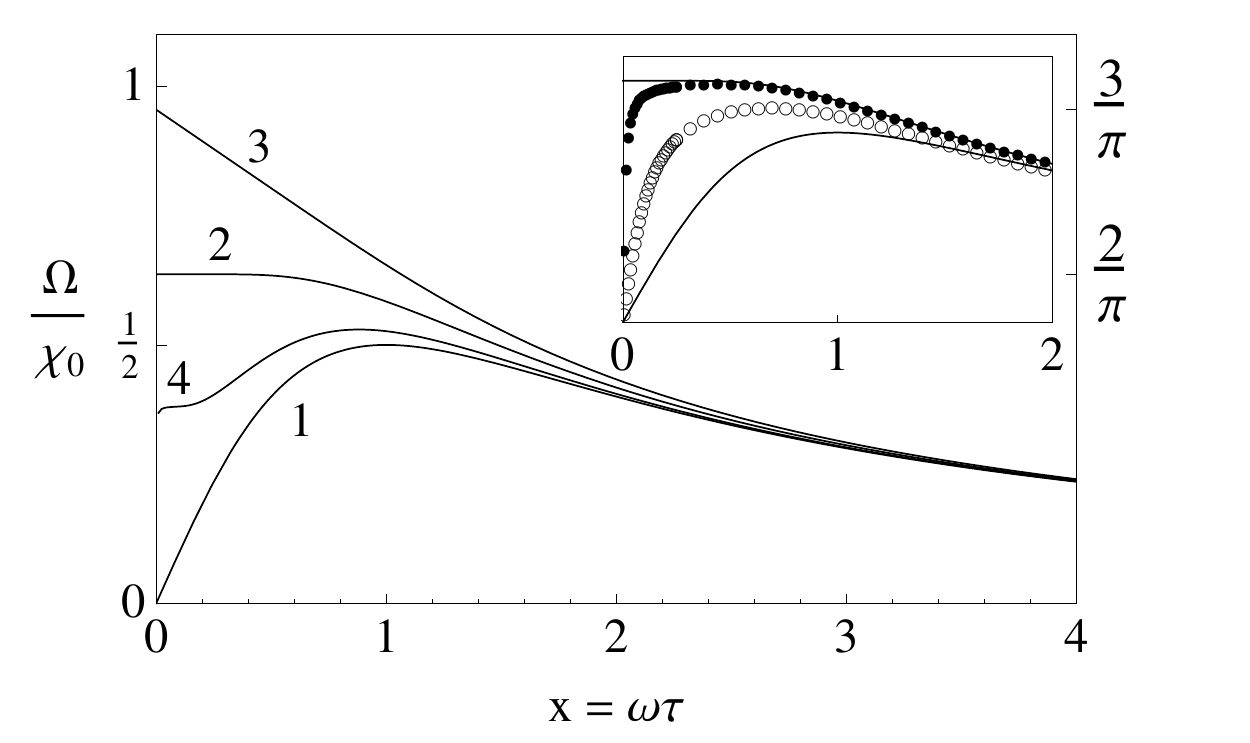}
\caption{\label{fig:sig1}Efficiency~vs.~$x$ for different signals. (curve 1) Harmonic wave, Eq.~(\ref{eq:effi_harm}); (curve 2) Square wave, Eq.~(\ref{eq:effi_square}); (curve 3) Sawtooth wave, Eq.~(\ref{eq:effi_sawtooth}); (curve 4) Step-ladder  with $\alpha=4$ [evaluated numerically from Eqs.~(\ref{eq:effi}) and (\ref{eq:eta_stepladder})]. (inset) Numerical evaluation of the sum in Eq.~(\ref{eq:effi}) for the square-wave extending up to $10$ (open circles) and $100$ (filled circles) harmonics. For comparative purposes, solid lines denote the harmonic and square-waves, as in the main plot. 
}
\end{figure}

As expected, for a harmonic signal [$\eta(t) = \cos t$], Eq.~(\ref{eq:effi}) yields:
\begin{equation}\label{eq:effi_harm}
\Omega(x) = \frac{x}{1+x^2} \quad \text{(harmonic wave)}
\end{equation}
This result is plotted in Fig.~\ref{fig:sig1}, curve 1, where it is seen to have a maxima at $x=\omega\tau =1$. 

One point requires further clarification: suppose we choose $\eta(t)$ in Eq.~(\ref{eq:effi}) such that $a_2=1$ and all other coefficients are zero (i.e., $\eta(t) = \cos(2t)$). This would then yield a result which is twice that obtained by replacing $x$ with $2x$ in Eq.~(\ref{eq:effi_harm}), a consequence of the $n^2$ term in Eq.~(\ref{eq:effi}). The reason for this apparent contradiction is that $\Omega(x)$ describes the energy conversion efficiency \emph{per period} for a signal whose \emph{fundamental period} is $2\pi$ (or $2\pi/\omega$ in real units). Hence the factor of two in this example would follow from counting the energy dissipated in two fundamental periods instead of one. A very important consequence follows from this argument: take, for concreteness, the point $x=1$  corresponding to the maxima of Eq.~(\ref{eq:effi_harm}). Since the corresponding maxima of higher order harmonics occur at different values of $x$, one may  argue that  replacing the main harmonic with a weighted sum containing higher harmonics  should always reduce the net dissipation. However, albeit having a smaller dissipation \emph{per fundamental period}, the higher harmonics are oscillating  with respect to the fundamental period of the first harmonic. In other words, the n-th order harmonic completes a total of $n$ periods during a time interval of $2\pi$,  thence compensating for it's lower dissipation.

%
%
In Fig.~\ref{fig:col} we summarize the signals investigated in this paper. We begin with the square wave [Fig.~\ref{fig:col} (a)]:
\begin{IEEEeqnarray}{rCl}
\label{eq:eta_square}
\eta (t) &=& 
\frac{1}{\sqrt{2}} \begin{cases} 
1	&	|t| \leq \frac{\pi}{2} \\
-1 	& 	\text{otherwise}
\end{cases} 
\end{IEEEeqnarray}
where the factor $1/\sqrt{2}$ was required in order to satisfy Eq.~(\ref{eq:eta_norm}).  Inserting the corresponding Fourier coefficients in Eq.~(\ref{eq:effi}), and using the partial fraction expansion of $\tanh(z)/z$, we obtain
\begin{equation}\label{eq:effi_square}
\Omega(x) = \frac{2}{\pi} \tanh\left(\frac{\pi}{2x}\right)\quad \text{(square wave)}
\end{equation}
This result is plotted in Fig.~\ref{fig:sig1}, curve 2. As can be seen, it is more efficient than the harmonic signal for all values of $x$ (i.e., it dissipates more energy for a given input energy). It also presents a ``plateau'' below $x=1$. This is important given that real systems always have some distribution of relaxation times. Thus, with a square wave the dissipation is expected to be more homogeneous. The fact that $\Omega(x\to0)\to 2/\pi$, which is clearly unphysical, is a consequence of the discontinuity of the input signal; evidently, in real systems $\Omega(x\to0)\to0$ since the input  is produced by a finite number of harmonics. To further emphasize this point we present on the inset of Fig.~\ref{fig:sig1} the result of numerically evaluating the sum [cf. Eq.~(\ref{eq:effi})] up to $10$ (open circles) and $100$ (filled circles) harmonics. As can be seen, increasing the number of harmonics significantly enhances the tendency of the curve to remain flat close to $x=0$.

Next, we turn to the sawtooth wave [Fig.~\ref{fig:col} (b)]:
\begin{IEEEeqnarray}{rCl}
\label{eq:eta_sawtooth}
\eta (t) &=& \frac{1}{\pi} \sqrt{\frac{3}{2}} t,
\end{IEEEeqnarray}
Inserting the corresponding Fourier coefficients in Eq.~(\ref{eq:effi}) and using, this time, the partial fractions expansion of $\coth(z)/z$, we obtain
\begin{equation}\label{eq:effi_sawtooth}
\Omega (x) = \frac{3}{\pi}\left[\coth\left(\frac{\pi}{x}\right) - \frac{x}{\pi}\right]\quad \text{(sawtooth wave)}
\end{equation}
This result is plotted in Fig.~\ref{fig:sig1}, curve 3. It dissipates more than both the harmonic and square waves, and has the zero-frequency limit $\Omega(x\to0)\to 3/\pi$, again due to the discontinuity. 

An important aspect of Eq.~(\ref{eq:effi}) is that the signal's shape enters only in terms of the combination $(a_n^2 + b_n^2)$. This means that entirely different signals may have the exact same efficiency. An example is the signal in Fig.~\ref{fig:col}~(c), corresponding to $\eta(t) = \log[2 \cos (t/2)] $. It's dissipative properties coincide exactly with those of the sawtooth wave. 

A closed form solution  for the (continuous) triangular wave also exists. Such, however, is of little practical interest since it  nearly coincides with the harmonic efficiency [cf. Eq.~(\ref{eq:effi_harm}) and Fig.~\ref{fig:sig1}, curve 1].  This is expected given the extremely rapid convergence of it's Fourier series. 

The step-ladder signal depicted in Fig.~\ref{fig:col} (d) approximately mimics real field variations. It can be written as 
\begin{IEEEeqnarray}{rCl}\label{eq:eta_stepladder}
\eta(t) &=& -1 + \frac{2}{\alpha-1} \sum\limits_{k=1}^{\alpha-1} \left[u\left(t+\frac{k\pi}{\alpha}\right) - u\left(t-\frac{k\pi}{\alpha}\right)\right]\!\!,\IEEEeqnarraynumspace
\end{IEEEeqnarray}
where $u$ is the unit-step function and $\alpha=2,3,4,\ldots$ defines the number of steps, with $2$ referring to the square-wave. The signal in Fig.~\ref{fig:col} (d) is for $\alpha=4$. 
In the limit $\alpha\to\infty$ we recover the harmonic field. A normalization constant $\sqrt{\frac{3(\alpha-1)}{2(\alpha+1)}}$ is also missing. 
No closed form solution exists for this signal. Notwithstanding, the corresponding sum may always be  evaluated numerically. Results for $\alpha=4$ are shown in Fig.~\ref{fig:sig1}, curve 4. As expected, it lies between the square and  harmonic waves, illustrating a gradual transition taking place between these two asymptotes.

%
%

Next we turn to pulsed signals. We begin with the rectangular pulse [Fig.\ref{fig:col}~(e)]:
\begin{IEEEeqnarray}{rCl}
\label{eq:eta_recpulse}
\eta (t) &=& 
\sqrt{\frac{\alpha}{2}} \begin{cases} 
1	&	|t| \leq \frac{\pi}{\alpha} \\
0 	& 	\text{otherwise}
\end{cases} 
\end{IEEEeqnarray}
Here $\alpha$ represent the width of the pulse: $\alpha=1$ correspond to a straight line and $\alpha\to\infty$ to a $\delta$-pulse. The function in Fig.~\ref{fig:col}~(e) is for $\alpha=4$ and the square-wave  [Eq.~(\ref{eq:eta_square})] correspond to $\alpha=2$. However, the signal is no longer symmetric with respect to $\eta=0$, which means that energy is being wasted in the  $a_0$ term of the Fourier series in Eq.~(\ref{eq:eta_fourier}). It thus follows that the efficiency for $\alpha=2$ will be half of that given by Eq.~(\ref{eq:effi_square}).  

The simplest way to obtain a general formula for the rectangular pulse [Eq.~(\ref{eq:eta_recpulse})], valid for all $\alpha>1$, is by induction.  For instance, when $\alpha=4$ the result is $\Omega_4(x) = (1/\pi)[\tanh(\pi/2x)+\tanh(\pi/4x)]$. The similarity between this result and Eq.~(\ref{eq:effi_square}) incites the idea that the induction formula may be written as a sum of hyperbolic tangents. Unfortunately, this is not the case.  However, if rewrite $\Omega_2$ and $\Omega_4$ in the form
\begin{IEEEeqnarray*}{rCl}
\Omega_2(x)&=&\frac{2}{\pi} \frac{\sinh(\pi/2x) \sinh(\pi/2x)}{\sinh(\pi/x)} \\[0.2cm]
\Omega_4(x)&=&\frac{4}{\pi} \frac{\sinh(\pi/4x) \sinh(3\pi/4x)}{\sinh(\pi/x)} 
\end{IEEEeqnarray*}
then another pattern becomes clearly visible inciting us to write: 
\begin{equation}\label{eq:effi_recpulse}
\Omega_\alpha(x) = \frac{\alpha}{\pi} \frac{\sinh\left(\frac{1}{\alpha}\frac{\pi}{x}\right)\sinh\left(\frac{\alpha-1}{\alpha}\frac{\pi}{x}\right)}{\sinh\left(\frac{\pi}{ x}\right)}
\end{equation}
(rectangular pulse). It turns out that this result is actually valid for \emph{all} values of $\alpha$  (with $\alpha>1$), including non-integers. Clearly, the correctness of the formula is easily tested by comparing it with the numerical calculation of the sum in Eq.~(\ref{eq:effi}).

\begin{figure}[!h]
\centering
\includegraphics[width=0.5\textwidth]{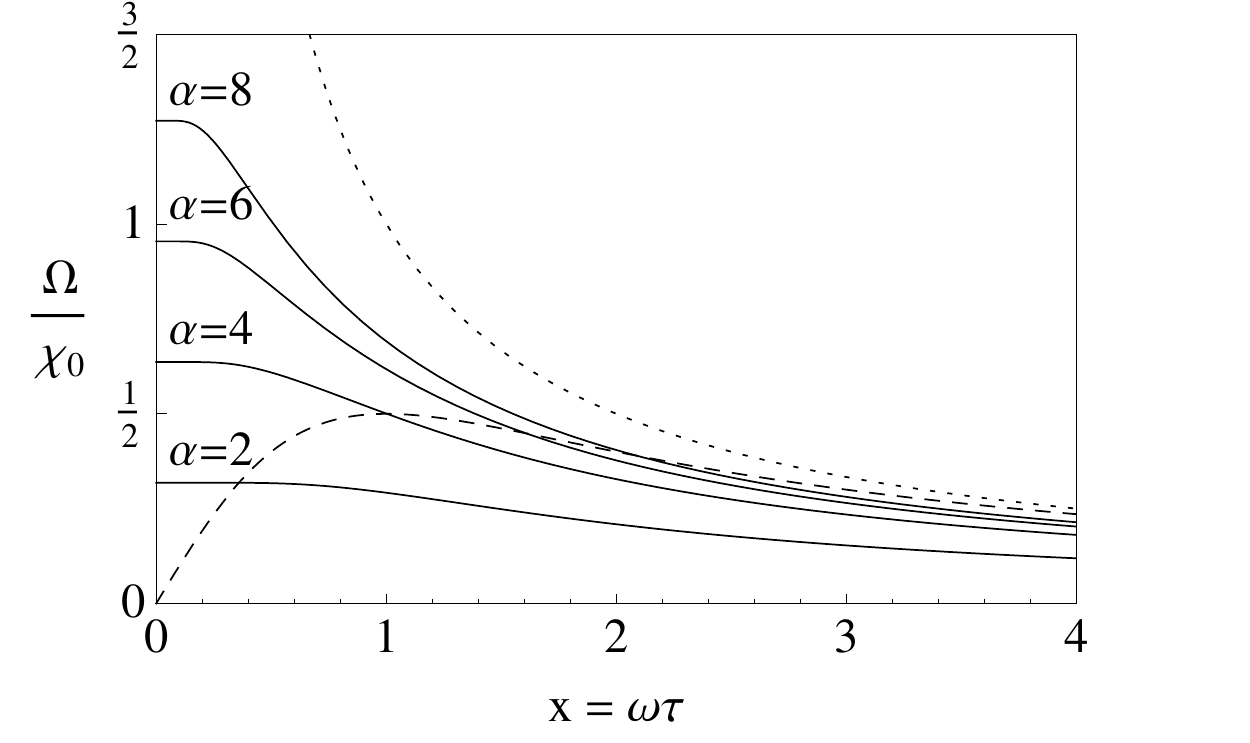}
\caption{\label{fig:sig2}Efficiency~vs.~$x$ for the rectangular pulse [Eq.~(\ref{eq:effi_recpulse})] for different values of $\alpha$. Harmonic efficiency is shown in dashed and $\delta$-impulse response, $\Omega = 1/x$, is shown in dotted.}
\end{figure}

Results for $\alpha =2$ (square wave), 4, 6 and 8 are shown in Fig.~\ref{fig:sig2} together with the harmonic response, shown in dashed for comparison. As can be seen,  the narrower the pulse (larger $\alpha$), the more efficient is the dissipation. For large $x$  the function behaves as $(1-(1/\alpha))/x$, which is smaller than the harmonic efficiency (which goes as $1/x$). At the other extreme, close to $x=0$, we have  that  $\Omega_\alpha(x\to0)\to \alpha/2\pi$; i.e., it scales linearly with $\alpha$.  Taking the limit $\alpha\to\infty$ we obtain the $\delta$-pulse response $\Omega_\infty (x) = 1/x$, which is illustrated in Fig.~\ref{fig:sig2} in a dotted line. To obtain the efficiency for a pulse symmetric with respect to $\eta=0$, one need only multiply Eq.~(\ref{eq:effi_recpulse}) by 2. In this case it is worth noting that for all pulses with $\alpha\geq2$, $\Omega_\alpha$  remains above the harmonic efficiency for all $x$.

Next we turn to the more realistic sine pulse [Fig.~\ref{fig:col}~(f)]:
\begin{IEEEeqnarray}{rCl}\label{eq:eta_sinepulse}
\label{eq:eta_sinp}
\eta (t) &=& \sqrt{\alpha}
 \begin{cases} 
\cos(\alpha t/2)	&	|t| \leq \frac{\pi}{\alpha} \\
0 	& 	\text{otherwise}
\end{cases}
\end{IEEEeqnarray}
The function in Fig.~\ref{fig:col}(f) is for $\alpha=4$. We restrict our analysis to $\alpha\geq1$, corresponding to the pulses with $\eta(\pm \pi) = 0$.
The calculations for this signal are somewhat more cumbersome. As before, the simplest approach is by induction. For conciseness, we give only the result:
\begin{IEEEeqnarray}{rCl}\label{eq:effi_sinp}
\Omega_\alpha (x) &=& \frac{\alpha^2 x}{[4+(\alpha x)^2]^2} \Bigg[4 + (\alpha x)^2-  \\[0.2cm] \nonumber
& &- \frac{8 \alpha x}{\pi} \frac{\cosh\left(\frac{\alpha-1}{\alpha}\frac{\pi}{x}\right)\cosh\left(\frac{1}{\alpha}\frac{\pi}{x}\right)}{\sinh\left(\frac{\pi}{x}\right)}\Bigg],
\end{IEEEeqnarray}
(sine pulse). Similarly to Eq.~(\ref{eq:effi_recpulse}), this formula turns out to be valid for all $\alpha>1$. 

\begin{figure}[!h]
\centering
\includegraphics[width=0.5\textwidth]{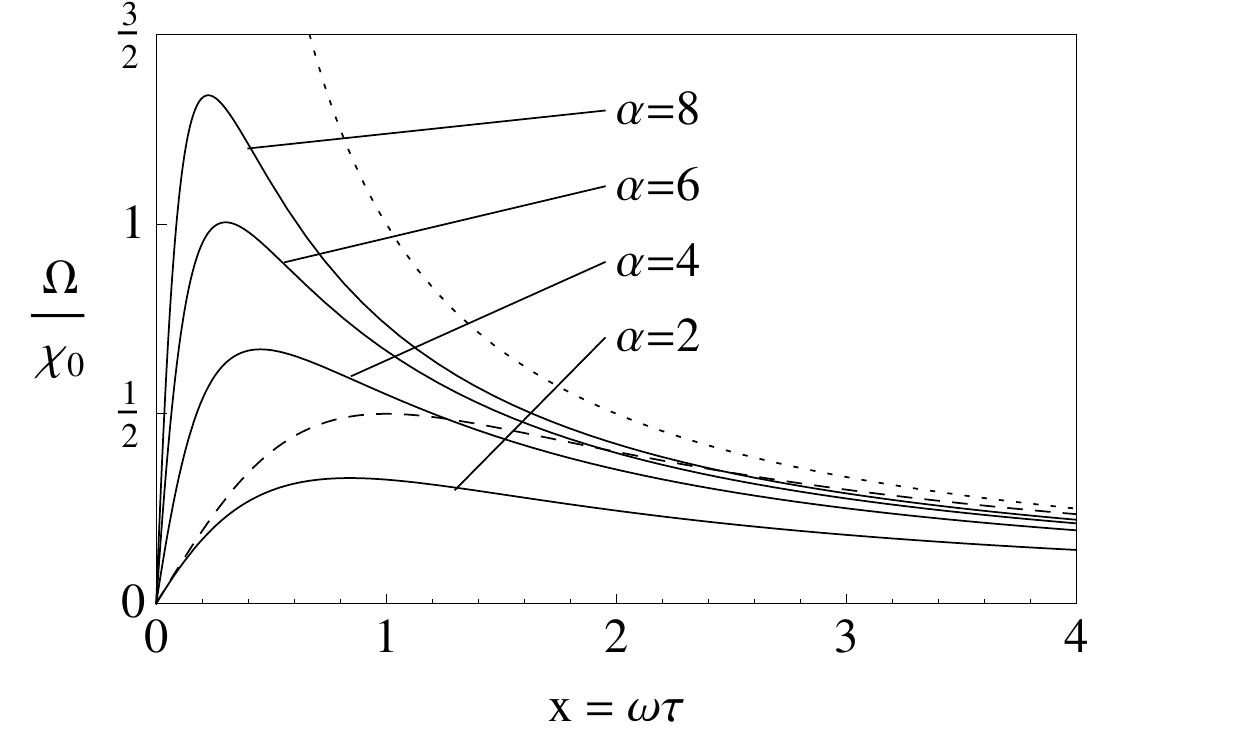}
\caption{\label{fig:sig3}Efficiency~vs.~$x$ for the since pulse [Eq.~(\ref{eq:effi_sinp})] for different values of $\alpha$. Harmonic efficiency is shown in dashed and $\delta$-impulse response, $\Omega = 1/x$, is shown in dotted.}
\end{figure}

Results for $\alpha=2$, 4, 6 and 8 are shown in Fig.~\ref{fig:sig3}, whose details are similar to Fig.~\ref{fig:sig2}. As can be seen, it shares a clear similarity with the rectangular pulse depicted in the latter. However, since it is continuous, we now have $\Omega(x\to0)\to0$. As before, sufficiently narrow pulses ($\alpha\gtrsim 4$) are considerably more efficient than the harmonic wave when $x<1$.  

%
%

We have thus far focused on an autocorrelation function of the form $C(t)= e^{-t/\tau}$. Near a critical point, however, the decay is known to become algebraic: $C(t) \propto t^{-\delta}$. For $\delta<1$ we may use Eq.~(\ref{eq:chi_kubo}) to show that $\Omega \propto \omega^\delta$ for all signal shapes; i.e., the only thing changing from one signal to another is the proportionally constant. Slight care must be taken, however, in the fact that the Fourier sum now reads $(a_n^2+b_n^2) n^{\delta+1}$ and may thus present convergence issues.

%
%

Finally, we apply these results to the problem of magnetic hyperthermia\cite{Carrey2011,Pankhurst2009}. The magnetic Langevin equation and it's corresponding Fokker-Planck equation were first introduced in Ref.~\onlinecite{Brown1963} and are described in detail in Ref.~\onlinecite{Coffey2004}. 
Conveniently, over a broad frequency interval (the ferromagnetic resonance region excerpted), the autocorrelation function  is adequately described by a single decaying exponential with relaxation time  $\tau(\sigma) = \frac{\tau_0}{2} \sqrt{\frac{\pi}{\sigma}} e^\sigma$. Here, $\tau_0\sim10^{-9}~\second$ and  $\sigma$ is the ratio between the energy  barrier separating the stable energy minima and the thermal energy\cite{Brown1963,Coffey2004}: $\sigma = Kv/k_B T$, where $K$ is the anisotropy constant and $v$ is the particle's volume.  On the other hand, the static susceptibility may be written as $\chi_0 \simeq \text{const} \times (\sigma-1)$. Finally, we have also fixed $\omega\tau_0 =10^{-4}$ for concreteness; the response to other frequencies is qualitatively similar. 

\begin{figure}[!h]
\centering
\includegraphics[width=0.5\textwidth]{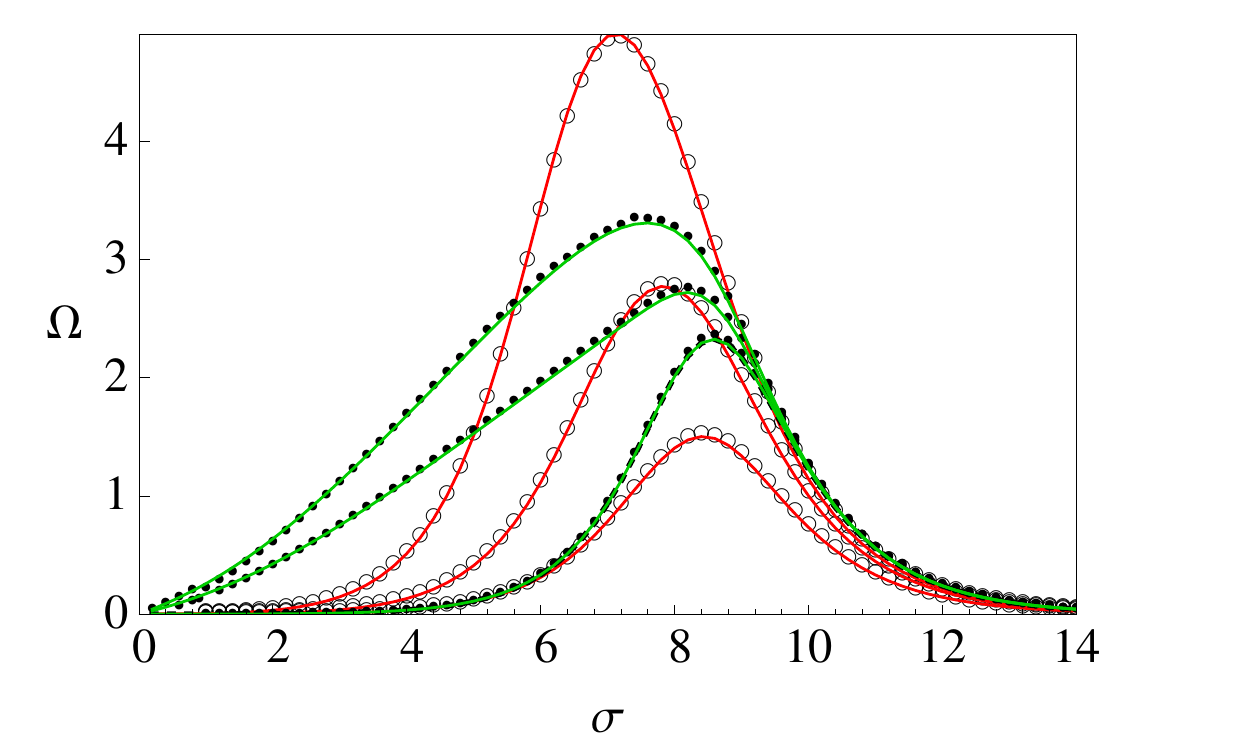}
\caption{\label{fig:sim}Efficiency~vs.~$\sigma$, the ratio of the energy barrier separating stable energy minima  to the thermal energy (see text for details). (green lines, open circles) Harmonic, square and sawtooth waves (from inner to outermost). (red lines, filled circles) Sine pulse with $\alpha=2$, 4 and 8 (from inner to outermost). Scattered points were computed from the Magnetic Langevin equation, simulating the magnetization $M(t)$ and numerically evaluating the area of the hysteresis loop. }
\end{figure}

In Fig.~\ref{fig:sim} we present in green lines (filled circles) the efficiency as a function of $\sigma$ for the harmonic, square and sawtooth waves (from inner to outermost). We also show in red  (empty circles) the calculations for the sine pulse [Eq.~(\ref{eq:effi_sinp})] with $\alpha=2$, 4 and 8 (again, from inner to outermost). Graphing $\Omega$~vs.~$\sigma$ enable us to directly related the efficiency to the fundamental magnetic parameter of the system. It can be seen that there is a maxima associated with each signal, corresponding to the value of $\sigma$ to which the particles should be tailored in order to maximize the dissipation. As for the asymptotic behavior of $\Omega$, when $\sigma$ is large all functions tend to zero exponentially. For small $\sigma$, on the other hand, the continuous signals (harmonic and sine pulse) scale roughly as $\Omega\propto \sigma^4$ whereas the discontinuous square and sawtooth waves present a linear dependence, $\Omega \propto \sigma$.

The scattered points in Fig.~\ref{fig:sim} were computed directly from the magnetic Langevin equation, first obtaining $M(t)$ from the numerical solution of a hierarchy of differential recurrence relations, and then the area [$A =\int\limits\!\!H\ud M$] by numerical integration. The computational details are described thoroughly in Ref.~\onlinecite{Landi2012b,*Landi2012e}. As can be seen, the agreement between both methods --- which are of \emph{entirely different nature} --- clearly establish the correctness of the formulas  presented in this paper. 

Finally, we acknowledge the  experimental challenges of producing non-harmonic AC magnetic fields  for  hyperthermia. The technique presently used is based on LC resonant circuits; whence, one alternative would be to stack synchronized circuits to produce the necessary harmonics. However, other alternatives may also exist for particular signals, such as the sawtooth wave which requires a ramp-like field increase, or general pulsed fields created from current pulses.  Another important topic regards the possible biological side effects of employing higher order harmonics (related to the formation of eddy currents). Unfortunately, we are at present unable to provide an adequate answer to this question in view of the lack of experimental results on the subject. We note, however, that the formation of eddy currents is proportional to the product $\omega H_0$. For this reason, the effect of higher order harmonics is expected to be, at least partially, mitigated by their smaller amplitudes (the Fourier coefficients always decay faster than $1/n$).

%
%

In conclusion, we have shown how Kubo's linear response theory may be extended to account for arbitrary signal shapes. The energy dissipated by the system was studied for several common signals and  analytical formulas were obtained in a variety of cases. It was shown that for the same energy input, substantial improvements in the dissipated output can be realized using different signal shapes. 
Even though the development was performed with respect to a magnetic system, the calculations here presented are entirely general and are expected to remain valid for any system where the linear response theory is applicable. By comparing the exact calculations with numerical simulations of magnetic hyperthermia we have (i) confirmed the exactness of our results and (ii) illustrated an important application of non-harmonic stimuli.

\acknowledgements
This work was supported by the Brazilian funding agency FAPESP.

%

\newpage 

\end{document}